# MeV electron acceleration at 1 kHz with <10mJ laser pulses

F. Salehi, A. J. Goers, G. A. Hine, L. Feder, D. Kuk, B. Miao, D. Woodbury, K. Y. Kim, and H.M. Milchberg

*Institute for Research in Electronics and Applied Physics*
*University of Maryland,*
*College Park, Maryland 20742, USA*

We demonstrate laser driven acceleration of electrons to MeV-scale energies at 1kHz repetition rate using <10mJ pulses focused on near-critical density He and $H_2$ gas jets. Using the $H_2$ gas jet, electron acceleration to ~0.5MeV in ~10fC bunches was observed with laser pulse energy as low as 1.3mJ. Increasing the pulse energy to 10mJ, we measure ~1pC charge bunches with >1MeV energy for both He and $H_2$ gas jets.

Laser-driven electron acceleration in plasma has become a well-established field since it was proposed several decades ago [1]. In recent years, significant experimental successes have been achieved, including the acceleration of quasi-monoenergetic electron bunches to ~4GeV [2] and the generation of MeV-range gamma rays [3]. Typically, these experiments demand laser pulse energies of at least several joules, and consequently existing laser technology limits them to low repetition rates (≤10Hz).

There are numerous applications for MeV-scale electron beams where a compact and portable high repetition rate source is beneficial, especially for potential scanning purposes and improved data collection statistics. At the low pulse repetition rates of ≤10Hz, radiography using laser-plasma-accelerated electron beams from gas jets [4,5], or γ-rays from bremsstrahlung conversion of the beam [6,7] has been demonstrated. Prior work at 0.5kHz using a continuous flow gas jet has produced ~100 keV electron bunches [8] and demonstrated their application to electron diffraction experiments [9]. While high repetition rate acceleration of electrons to MeV-scale using solid and liquid targets has been reported [10,11], gas jet-based laser-plasma electron sources had yet to simultaneously achieve high repetition rate and MeV-scale energies.

In non-plasma based work, ultrafast electron diffraction using laser-driven photocathodes and conventional accelerator structures such as LINACs is an established research area [12], but it is difficult to achieve <100fs temporal resolution with such electron pulses due to timing jitter and space charge effects.

The most common and successful laser-plasma-based acceleration scheme is laser wakefield acceleration (LWFA), which can be initiated by relativistic self-focusing of the laser pulse in the plasma. LWFA electron pulses can be ultrashort and are precisely timed to their driving pulses [13]. Relativistic self-focusing has a critical power [14] of $P_{cr}$=17.4($N_{cr}/N_e$) GW, where $N_e$ is the plasma density and $N_{cr}$ is the critical density. As $N_{cr}$=1.74×10$^{21}$cm$^{-3}$ for the Ti:Sapphire laser wavelength of λ=800nm, a very high $N_e$ is needed to keep $P_{cr}$ well below 1 TW and enable operation with current commercial laser technology for millijoule-scale pulses at 1 kHz. In previous experiments, we showed that the use of a high density gas jet (at $N_e/N_{cr}$<0.25) lowers $P_{cr}$ sufficiently to promote relativistic self-focusing and acceleration in the self-modulated laser wakefield (SM-LWF) regime with subterawatt laser pulses [15]. In this Letter, we show that using gas jets approaching even closer to critical density makes possible electron acceleration to relativistic energies with pulse energies as low as 1.3mJ, delivered at 1kHz. We note that for pulse propagation near $N_e/N_{cr}$ ~ 0.25, the stimulated Raman scattering associated with SM-LWF generation can compete with the two-plasmon decay instability [16]. To help understand the details of laser propagation and acceleration in this regime, we present particle-in-cell (PIC) simulations later in this paper.

Driving laser plasma accelerators at high repetition rate demands an interaction target with a high duty cycle. For a gas jet target, this means a nearly continuous flow of gas out of a high pressure nozzle, and significant accumulated gas loading of the experimental vacuum chamber. Such gas loading leads to high background chamber pressure, which can enhance the deleterious effects of laser-induced ionization and defocusing well before the pulse encounters the gas jet. Our experiments demonstrate electron acceleration at chamber background pressures as high as 20 Torr, enabling use of continuous flow nozzles and even higher repetition rate laser systems for LWFA.

and $H_2$ jets. We used λ=800nm, 30fs, <12mJ pulses from a 1kHz Ti:Sapphire laser to drive LWFA in the dense jets. The pulses were tightly focused with an f/8.5 off-axis paraboloid to a ~ 9µm intensity FWHM spot size. Pulse energy was controlled with a half-wave plate before the compressor gratings, enabling energy scans in the range ~0.1mJ – 12mJ by rotation of the laser pulse's linear polarization with respect to the grating rulings. Given the risk of high accumulated gamma radiation dose from running the experiment at 1kHz (mainly from the beam dump), we used a solenoid valve before the nozzle to control the gas flow duration from less than 1ms up to several minutes. The electron beams in Fig. 1 were generated by 9.5mJ laser pulses with a He gas jet open time of 20ms.

Gas jet density and plasma profiles were measured using folded wavefront interferometry [17] with a λ=800nm probe split from the main pulse. High density $H_2$ and He gas jets were produced by cooling the gas to −150C while pressurized up to 1100psi, and flowing the gas through a 150µm internal diameter nozzle into a vacuum chamber pumped by a 220CFM roots blower. The gas jet density encountered by the laser pulse was controlled by changing the backing pressure, temperature, and the location of the laser focus on the jet. As determined from interferometry, the jet density has a Gaussian transverse profile of FWHM ~150-250µm depending on laser focus position with respect to the nozzle orifice. Within ~60 µm of the nozzle exit, we achieve $N_e/N_{cr}$~1 at full ionization. To avoid nozzle damage, the laser was focused at least ~110µm above the nozzle orifice, where $N_e/N_{cr}$~0.5. Accelerated electron spectra were collected 35cm beyond the jet by a magnetic spectrometer consisting of a compact permanent 0.08T magnet located behind a 1.7mm wide copper slit, followed by a LANEX scintillating screen imaged onto low noise CCD camera. The LANEX screen was shielded from laser light by 25µm thick aluminum foil. Accelerated electron beam profiles were collected by moving the slit and magnet out of the way. A lead brick electron beam dump was placed behind the LANEX and turning mirror. Day-to-day experimental runs for similar jet opening times gave varying electron bunch energies and charges owing to gas jet nozzle tip erosion from plasma ablation. Nozzles were replaced after approximately 2×10⁵ laser shots.

Figure 2 shows accelerated electron spectra from the $H_2$ jet for several values of laser pulse energy and with 10ms valve open time. The inset shows the total charge per shot accelerated to >1MeV energy vs. laser pulse energy. Each point is the average of 10 consecutive shots. Lowering the laser pulse energy requires increasing the electron density (via the jet gas density) to maintain $P>P_{cr}$. The minimum electron density required to observe electron acceleration with 9mJ pulses was ~$4.0 \times 10^{20}$ $cm^{-3}$ ($N_e/N_{cr}$= 0.23). To observe acceleration for 1.3mJ pulses, it was necessary to increase the electron density to ~$1.2 \times 10^{21}$ $cm^{-3}$ ($N_e/N_{cr}$= 0.69).

At low laser pulse energies (< 3mJ) with $H_2$ jets, most of the electrons are at energies below our spectrometer range and are excessively deflected by the magnet. Moving the spectrometer out of the electron beam path allows the full beam to impact the LANEX (shielded by 25µm aluminum foil). Using the electron transmission data for aluminum [18] and the LANEX response [19,20], we estimate electron bunches of ~10 fC charge with up to ~0.5MeV energy for laser pulse energies as low as 1.3mJ.

For He jets no electron beams were detected for laser pulses <5mJ. For both $H_2$ and He jets, increasing the pulse energy to

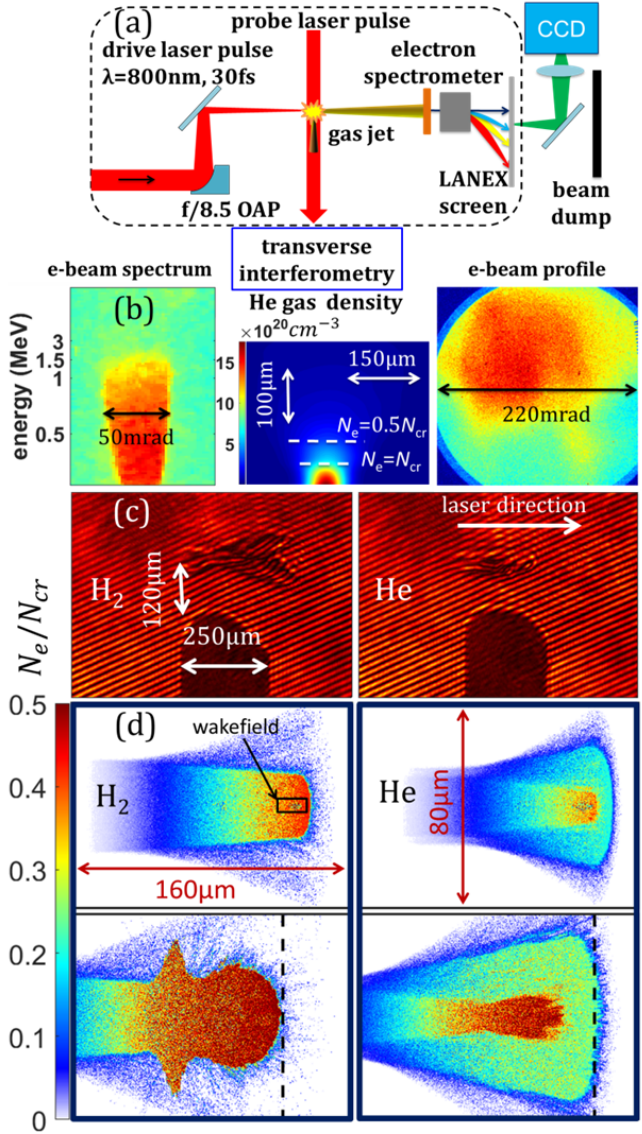

**Fig. 1.** (a) Experimental setup for high repetition rate electron acceleration. The dashed line depicts the vacuum chamber boundary. (b) Measured density profile for He gas jet (center), electron beam profile from 20 consecutive shots at 1kHz with 9.5mJ pulse energy on He jet (right), and corresponding electron energy spectrum (left). The sharp left-right edges on the spectrum are from electron beam clipping on the spectrometer magnet, and the lower energy section is focused by the magnet's fringe fields. (c) Interferograms showing residual plasma ~1ps after interaction of 5 mJ pulses with $H_2$ and He gas jets. The dark shadow is the gas nozzle. (d) Electron density profiles before (top) and 250fs after wavebreaking (bottom) from 2D PIC simulations of interaction of 5 mJ, 30fs laser pulses with 200 µm FWHM $H_2$ and He jets at peak neutral density 4.35 ×10²⁰ molecules or atoms per cm³. The dashed vertical lines indicate the centre of the gas jet.

Figure 1 shows the experimental setup along with a measured He gas density profile, an accelerated electron beam profile, corresponding electron energy spectra, and interferograms and simulation results showing laser-generated plasma in the He

~10mJ increased the bunch charge with >1MeV energy to ~ 1pC. We attribute these observations to ionization-induced defocusing in He at low laser pulse energy. The transverse electron density profile in the $H_2$ jet is flatter than in the He jet owing to lower threshold for full ionization in $H_2$ [21], resulting in less defocusing in $H_2$ and larger amplitude plasma waves This is borne out by interferograms (Fig. 1(c)) showing the residual plasma ~1 ps after interaction of a 5mJ pulse with the He and $H_2$ jets. The associated 2D PIC simulations (Fig. 1(d)) using the code TurboWave [22] show the electron density profiles just before and 250fs after plasma wavebreaking in the $H_2$ and He jets—it is seen that the hydrogen plasma profile is fully ionized over a wider region than in He, and that post-wavebreaking scatter of the laser pulse and electron heating in hydrogen gives a wider profile at the jet exit.

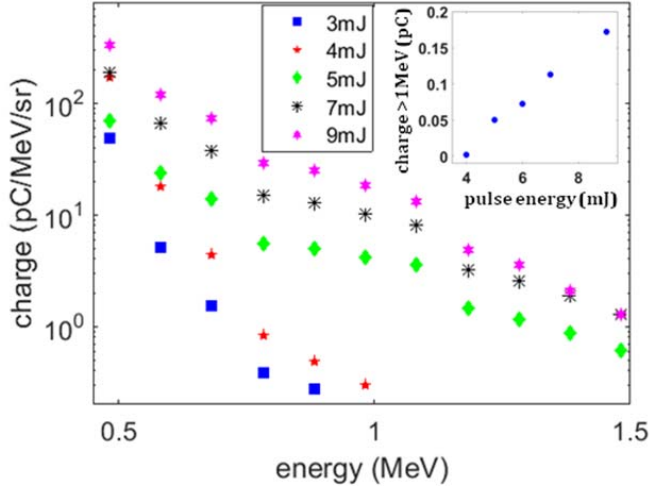

**Fig. 2.** Accelerated electron energy spectra from $H_2$ jets for varying laser pulse energy and 10ms gas jet open time. The inset shows total charge with > 1MeV energy vs. laser pulse energy.

Figure 3a shows results from the He jet using 9.5mJ pulses and a 20ms valve open time, with accelerated electron spectra for varying peak electron density and corresponding total charge accelerated to >1MeV in the inset. Figure 3b shows electron beam profiles on LANEX for selected He plasma densities of Fig. 3a, showing the sensitivity to plasma density. While the total accelerated charge increases significantly with peak electron density, the normalized electron spectrum does not change noticeably. The beam divergence angle (estimated from an average around the 50% beam intensity contour) is ~150mrad at $N_e/N_{cr}$=0.25 and increases to ~260mrad as the electron density is increased to $N_e/N_{cr}$=0.43.

A major concern using a high density continuous flow gas jet is the background pressure buildup inside the target chamber, which can prevent the laser pulse from interacting with the highest density part of the jet at the highest intensity owing to ionization-induced defocusing of the pulse. In order to study the effect of background pressure buildup, we first measured accelerated electron spectra for increasing valve open times (with the laser at 1 kHz and the jet repetition rate at 0.5Hz), as shown in Figure 4, where a He gas jet at $N_e/N_{cr}$=0.54 is driven by 10mJ laser pulses. It is seen that increasing the valve open time lowers the charge per shot while keeping the normalized spectra similar, with the charge at > 1MeV decreasing from ~1.6pC to ~0.2pC over the opening times 1ms-100ms, over which the corresponding background pressure increased from < 0.1Torr to ~ 3.5Torr.

Increasing the valve open time to 1sec, with a repetition rate of 0.5Hz, increases the background pressure to a constant ~20Torr. Scanning a 50ms window (containing a 50 shot burst of 10mJ pulses) over the 1 sec valve opening of the He jet gives a nearly unchanging LANEX signal. This shows that the valve could be open continuously if the accumulation of gamma ray dose from our beam stop was not a constraint.

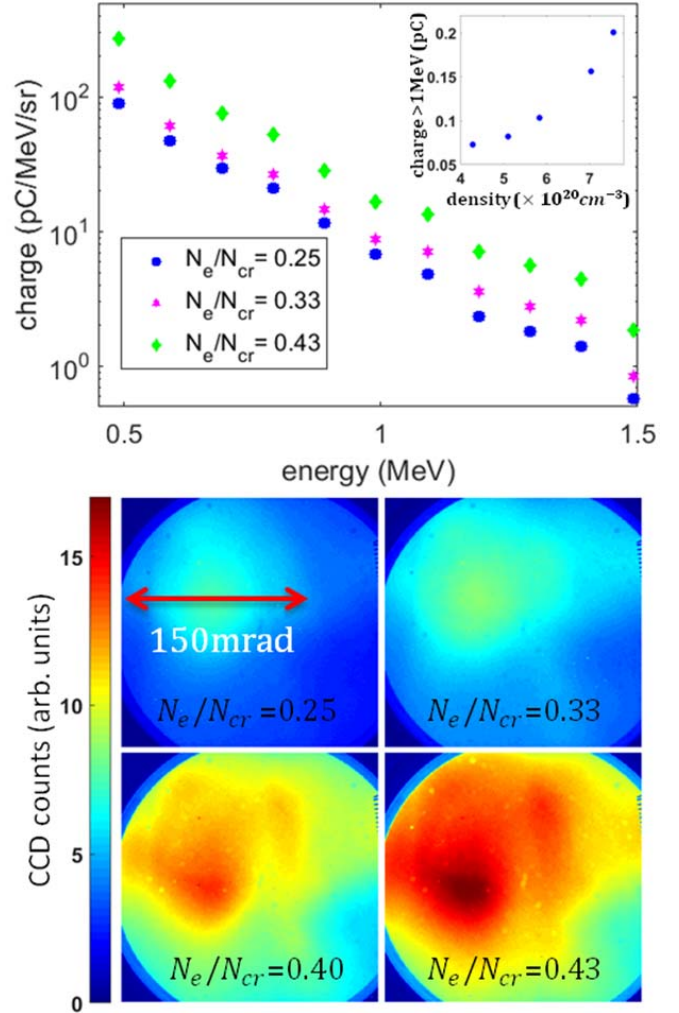

**Fig. 3.** *Top*: Electron energy spectrum for varying plasma density from He jet using 9.5mJ laser pulses and 20ms gas jet open time. The inset shows total charge per shot with >1MeV energy. *Bottom*: Electron beam profiles on LANEX screen, illustrating sensitivity to plasma density. The outside circle is the outline of the vacuum port, through which the LANEX surface was imaged.

To better understand SM-LWF generation and acceleration in our jet at electron densities above quarter critical ($N_e/N_{cr}$>0.25), we performed 2D PIC simulations for 4mJ laser pulses interacting with a 200μm FWHM preionized $H_2$ target with peak $N_e/N_{cr}$ = 0.5. Figure 5 shows the simulated plasma wake just before and after wavebreaking (top) and corresponding central lineouts (bottom) of density and normalized laser vector potential $a_0$. The wakefield is generated at ambient plasma density above quarter critical (dashed line), where the Raman Stokes line is suppressed and the anti-Stokes line dominates, as seen in the spectrum shown. Two-plasmon decay is not evident over the full laser propagation, possibly due to the strongly nonlinearly steepened density in the plasma wake [23]. These considerations are an important element of the dense jet interaction physics being studied in our ongoing experiments and simulations.

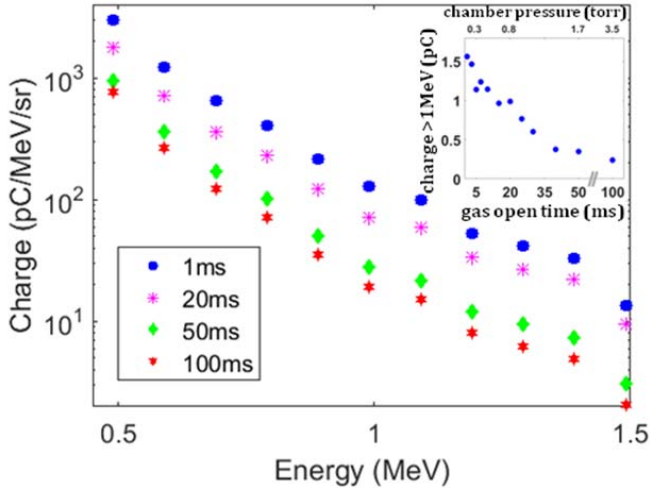

**Fig. 4.** Electron energy spectrum from He gas jet ($N_e/N_{cr}$=0.54) for different valve open times for 10mJ laser pulses. Inset: Total charge per shot accelerated to > 1MeV and corresponding background pressure.

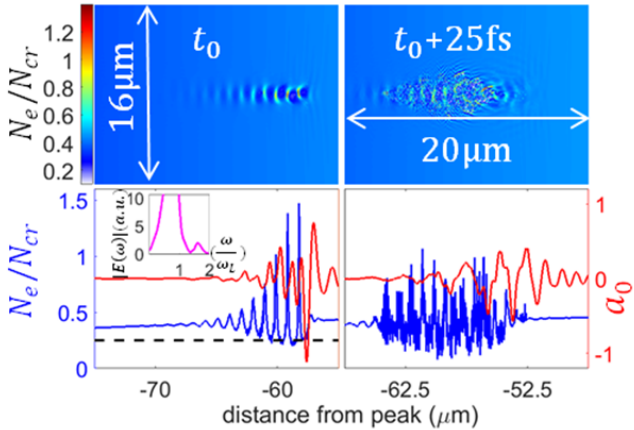

**Fig. 5.** Simulated plasma wake just before and after wavebreaking (top) and corresponding central lineouts (bottom) of density and normalized laser vector potential, for 4mJ pulse interacting with 200μm FWHM preionized $H_2$ target of peak $N_e/N_{cr}$ = 0.5. Dashed line: $N_e/N_{cr}$ = 0.25. Inset: Pre-wavebreaking spectrum of self-modulated laser showing anti-Stokes line, with Stokes line suppressed.

In summary, we have demonstrated for the first time, to our knowledge, laser driven electron acceleration to >1MeV in a gas jet using a 1kHz repetition rate mJ-scale laser, with bunch charge to the pC level. This result was made possible by use of a thin, dense, gas jet target enabling near-critical density laser interaction. Such a high repetition rate, high flux ultrafast source has immediate application to time resolved probing of matter for scientific, medical, or security applications, either using the electrons directly or using a high-Z foil converter to generate ultrafast γ-rays.

This work was supported by the U.S. Department of Energy (Grant Nos. DESC0015516 and DESC0010706TDD), the Air Force Office of Scientific Research (Grant No. FA95501310044), the National Science Foundation (Grant No. PHY1535519), and the Department of Homeland Security (Grant No. 2016DN077ARI104). The authors thank Yungjun Yoo for help with the laser.